\documentclass{PoS}

\usepackage{amsmath}
\usepackage{graphicx}

\newlength{\dummysp}
\newcommand{\bbar}[1]{{\overline{#1}}}
\newcommand{\half}{{\frac{1}{2}}}

\newcommand{\beq}{\begin{eqnarray}}
\newcommand{\eeq}{\end{eqnarray}}
\newcommand{\nnn}{ \nonumber \\ }
\newcommand{\p}{{\partial}}

\newcommand{\s}{{\sigma}}
\newcommand{\vev}[1]{{\langle #1 \rangle}}

\newcommand{\gappeq}{\mathrel{\rlap {\raise.5ex\hbox{$>$}}
{\lower.5ex\hbox{$\sim$}}}}
\newcommand{\lappeq}{\mathrel{\rlap{\raise.5ex\hbox{$<$}}
{\lower.5ex\hbox{$\sim$}}}}
\newcommand{\myref}[1]{(\ref{#1})}

\newcommand{\bfe}[1]{\vspace{5pt} {\bf #1 \hspace{2pt}}}

\newcommand{\ben}{\begin{enumerate}}
\newcommand{\een}{\end{enumerate}}

\newcommand{\sbar}{{\bar \s}}
\newcommand{\phib}{{\bar \phi}}

\newcommand{\psib}{{\bar \psi}}

\newcommand{\bit}{\begin{itemize}}
\newcommand{\eit}{\end{itemize}}

\newcommand{\Ncal}{{\cal N}}

\newcommand{\zb}{{\bbar{z}}}

\newcommand{\Fb}{{\bar F}}

\newcommand{\Qb}{{\bar Q}}

\newcommand{\Wbar}{{\bbar{W}}}
\newcommand{\rsp}{\cite{Giedt:2005ae}}

\def\[{\left [}
\def\]{\right ]}
\def\({\left (}
\def\){\right )}

\author{\speaker{Joel Giedt}\thanks{This
work was supported by the National Science and Engineering 
Research Council of Canada and the Ontario 
Premier's Research Excellence Award.} \\
                                                                                
        Department of Physics, University of Toronto \\

		  60 St. George St., Toronto, ON,  M5S 1A7, Canada \\
                                                                                
        E-mail: \email{giedt@physics.utoronto.ca}}

\title{Symmetry and scaling in the Q-exact
lattice (2,2) 2d Wess-Zumino model}
                                                                                
\ShortTitle{Symmetry and scaling in the Q-exact
lattice (2,2) 2d Wess-Zumino model}

\abstract{As a nonperturbative check on the Q-exact lattice 
formulation, we demonstrate that the continuum R-symmetries
are recovered.  We locate the critical domain of the
lattice theory.  Aspects of the continuum nonrenormalization theorems
are found to be respected at finite lattice spacing.
Preliminary attempts to extract
critical exponents---another nonperturbative 
check---are discussed.  All of our results are
obtained from Monte Carlo simulations with
dynamical fermions.}

\FullConference{XXIIIrd International Symposium on Lattice Field Theory\\
                                                                                
                 25-30 July 2005\\
                                                                                
                 Trinity College, Dublin, Ireland}

\PoS{PoS(LAT2005)270}

\begin{document}

\bfe{Motivation.}
The continuum (2,2) 2d Wess-Zumino (2dWZ) model
(obtained from a dimensional reduction
of the 4d Wess-Zumino model \cite{Wess:1973kz})
is supposed to provide a Landau-Ginzburg
description of the minimal discrete series
of $\Ncal=2$ superconformal field theories~\cite{Boucher:1986bh}.
In our recent article \rsp, 
we have examined an important
aspect of the simplest of these models---the 
one with a cubic superpotential---in
the context of a class of lattice
actions that have an exact lattice 
supersymmetry.
These lattice actions were first formulated 
in \cite{Elitzur:1983nj,Sakai:1983dg} using 
{\it Nicolai map} \cite{Nicolai:1979nr} methods, 
relying on earlier Hamiltonian \cite{Elitzur:1982vh}
and continuum \cite{Cecotti:1982ad} studies
that also utilized the Nicolai map.
Detailed studies of the spacetime lattice system
were performed in \cite{Beccaria:1998vi} by stochastic
quantization methods and in \cite{Catterall:2001wx}
by the Monte Carlo simulation approach.

Once auxiliary fields are introduced, the
lattice action takes a Q-exact form: $S=QX$,
as was emphasized in the topological
interpretation of \cite{Catterall:2003xx}
and the lattice superfield approach of \cite{Giedt:2004qs}.  
Here $Q$ is a lattice supercharge with
derivatives realized through discrete difference
operators; with respect to a
discrete approximation of the continuum
theory superalgebra, $Q^2=0$ is a nilpotent
subalgebra.  Because $S$ is Q-exact, the action is
trivially invariant with respect to this
lattice supersymmetry:  $QS=Q^2X=0$.

It was shown in \rsp\ that in the massive continuum theory
there is an exact $Z_2(R)$ symmetry.  It is an {\it R-symmetry,}
meaning it does not commute with the supercharges.
This symmetry is spontaneously broken at infinite
volume.  In the massless case, i.e., in the critical
domain, the classical R-symmetry
is enlarged to $U(1)_R$.  It cannot be spontaneously
broken since it is a continuous symmetry in 2d~\cite{MWC}.
If the lattice theory has the correct continuum
limit, it should reproduce these features.  On
the other hand, these R-symmetries are only
approximate in the Q-exact lattice action; the
symmetry is explicitly broken by the Wilson
mass term that is used to lift doublers.\footnote{This
is directly related to the breaking of
the so-called $U(1)_V$ symmetry, that was
pointed out in \cite{Giedt:2004qs}.}

It has been shown in \cite{Giedt:2004qs} that 
the continuum limit
of the lattice perturbation series is identical
to that of the continuum theory,
due to cancellations that follow from $Q^2=0$.
Thus, the Q-exact spacetime lattice has
behavior that is similar to what was
found on the $Q,Q^\dagger$-preserving
spatial lattice in \cite{Elitzur:1983nj}.
However, it was also shown in \cite{Giedt:2004qs}
that the most general continuum
effective action that is consistent
with the symmetries of the bare lattice action
is not the (2,2) 2d Wess-Zumino model.
This raises the question of whether or not the good behavior
of perturbation theory persists at a nonperturbative level.
The results of \cite{Beccaria:1998vi,Catterall:2001wx}
give hope that the desired continuum limit
is obtained beyond perturbation theory.  
If so, this would be one of the few
examples of a supersymmetric field theory that
can be latticized and studied nonperturbatively
by Monte Carlo simulation
without the need for fine-tuning
of counterterms.

In our recent work \rsp, we have shown that
features of the continuum theory associated with
the R-symmetry are recovered in the continuum
limit; this provides further 
evidence that the correct theory is
obtained.  The symmetry that we study
persists in the infrared effective theory
in a strongly coupled regime.  Thus,
we are testing aspects
of the lattice theory that lie beyond
the reach of perturbation theory.
In the remainder of this note,
we summarize the main results of \rsp.

\bfe{Symmetries of the continuum.}
The continuum Euclidean action is
\beq
S &=& \int d^2z \; \[ - 4 \phib \p_z \p_\zb \phi 
-2i \psib_- \p_\zb \psi_- + 2i \psi_+ \p_z \psib_+ 
- \Fb F \right. \nnn
&& \left. + W'(\phi) F + \Wbar'(\phib) \Fb
- W''(\phi) \psi_+ \psi_- - \Wbar''(\phib) \psib_- \psib_+ \]
\label{cact}
\eeq
We specialize to the superpotential
$W(\phi) = \frac{m}{2} \phi^2 + \frac{g}{3!} \phi^3$.
It is convenient to make the field redefinition
$\phi = -\frac{m}{g} + \s$.
In this case $W(\s) = - \lambda \s + \frac{g}{3!} \s^3, \;
\lambda = \frac{m^2}{2g}$.
The scalar potential is just
$V = |W'(\s)|^2 = \left| \lambda - \frac{g}{2} \s^2 \right|^2$.
Degenerate minima occur:
$\s_\pm = \pm \sqrt{2\lambda/g} =\pm m/g$.
The action possesses a (2,2) supersymmetry,
characterized by the algebra
$\{ Q_- , \Qb_- \} = -2i \p_z, \;
\{ Q_+ , \Qb_+ \} = 2i \p_\zb$.
In the case of $\lambda=0$, there is a $U(1)_R$
symmetry: $\s \to e^{2 i \alpha /3} \s, \;
\psi_\pm \to e^{-i \alpha/3} \psi_\pm, \;
F \to e^{-4i \alpha/3} F$.
In addition there is a $U(1)_A$ symmetry:
$\psi_\pm \to e^{\pm i \omega} \psi_\pm, \;
\psib_\pm \to e^{\mp i \omega} \psi_\pm$,
with all other fields neutral.
If $\lambda \not=0$, the symmetry breaking
$U(1)_R \times U(1)_A \to Z_2(R) \times U(1)_A$
occurs, with $Z_2(R)$ described by:
$\s \to - \s, \; \sbar \to - \sbar, 
\; \psi_\pm \to \pm \psi_\pm,
\; \psib_\pm \to \pm \psib_\pm$.
An important property of the theory
is that there is only wavefunction
renormalization: $W(m,g|\phi) \equiv W(m_r,g_r|\phi_r)$,
where $m_r = Z m, \; g_r = Z^{3/2} g, \;
\lambda_r = \sqrt{Z} \lambda$.
This is the so-called nonrenormalization
theorem:  mass and coupling counterterms
vanish identically.
It follows that $m=0$ is a critical point for
any $g$.

\bfe{Q-exact lattice action and its symmetries.}
The lattice action preserves the nilpotent subalgebra
$Q^2 = 0$ where $Q=Q_- + \Qb_+$.
The action of $Q$ on lattice fields is defined by
a discretized version of the continuum supersymmetry:\footnote{
Here and below
the difference operators are defined as
$\Delta_\mu^S = \half \( \Delta^+_\mu + \Delta^-_\mu \), \;
\Delta^2 = \sum_{\mu=1,2} \Delta^+_\mu \Delta^-_\mu, \;
\Delta_z = \half \( \Delta^S_1 - i \Delta^S_2 \), \;
\Delta_\zb = \half \( \Delta^S_1 + i \Delta^S_2 \)$,
where $\Delta_\mu^+$ and $\Delta_\mu^-$ are forward and
backward difference operators in the $\mu$ direction.}
$Q \phi = \psi_-, \;
Q \psi_+ = F + 2i \Delta_\zb \phi, \;
Q \psi_- = 0 , \;
Q F = -2i \Delta_\zb \psi_-, \;
Q \phib = \psib_+ , \;
Q \psib_+ = 0 , \;
Q \psib_- = \Fb - 2i \Delta_z \phib , \;
Q \Fb = 2i \Delta_z \psib_+$.
The action is Q-exact:
\beq
S &=&  Q \( -F \psib_{-} 
- 2i \psi_{+} \Delta_{z} \phib
+ W'(\phi) \psi_{+} + \bbar{W}'(\phib) 
\psib_{-} \) 
\label{iio}
\eeq
The auxiliary fields $F,\Fb$ can be eliminated
by their equation of motion.
To lift spectrum doublers, a Wilson mass term
is introduced into the superpotential:
\beq
W(\phi) =  \sum_m \( - \frac{r}{4} \phi_m \Delta^2 \phi_m
+ \frac{m}{2} \phi_m^2 + \frac{g}{3!} \phi^3_m \)
\eeq
Note that $W'_m = \p W/\p \phi_m$, etc.
We can also make use of the $\phi \to \s$
field redefinition to obtain
\beq
W(\s) =  \sum_m \( - \frac{r}{4} \s_m \Delta^2 \s_m
- \lambda \s_m + \frac{g}{3!} \s^3_m \)
\eeq
Note that the Wilson mass term (an irrelevant operator)
violates the R-symmetries
of the continuum theory.  In \rsp\
we have shown that the effective potential
nevertheless has the continuum symmetries:
the effect of the irrelevant symmetry breaking operator
is negligible for small lattice spacing.

We probe the symmetry of the effective
potential by introducing a background
field in the scalar potential:
$\Delta V(h) = - \sum_m \( \bar h \s_m - h \sbar_m \)$.
This allows us to explore the extent to
which the lattice theory is symmetric w.r.t.~$\s \to -\s$,
or the phase rotation $\s \to e^{i\theta} \s$.
Define the generating function $w(h)=\ln Z(h)$, where
$Z(h)$ is the partition function that is
obtained when $\Delta V(h)$ is added to the lattice
action.  $Z_2(R)$ symmetry of the effective
potential is equivalent to $w(-h)=w(h)$.
Similarly, $U(1)_R$ symmetry of the effective potential
is equivalent to $w(e^{i \theta} h) = w(h)$.
Note also that $\vev{\s}_h = \p w(h) / \p \bar h$,
where $\vev{\s}_h$ is the expectation value
of $\s$ in the background $h$.
It follows that in the case of $Z_2(R)$ symmetry
we have the prediction
$\vev{\s}_{-h} = - \vev{\s}_h$.
In the case of $U(1)_R$ symmetry we have
the much stronger prediction
$\vev{\s}_{e^{i \theta} h} = e^{i \theta} \vev{\s}_h$.
Equivalently, since we take $m>0,g>0$ and will
find below that $\vev{\s}_h > 0$ if $h$ is real and positive,
\beq
\arg \vev{\s}_h = \arg h, \quad {\rm and} \quad
|\vev{\s}_h| = {\rm const.,} \quad {\rm fixed} ~ |h|
\label{u1con}
\eeq

In \rsp\ we have shown by various means
that, up to statistical errors, simulation
results are supportive of
the $Z_2(R)$ symmetry prediction
$\vev{\s}_{-h} = - \vev{\s}_h$.
We have also shown that $m=0$ is a critical point.
For brevity we do not discuss the details here.
Rather, we will concentrate on the $U(1)_R$
symmetry at $m \to 0$; however, some of the
results we review here also indicate the
$Z_2(R)$ symmetry at $m \not= 0$.

\begin{figure}
\begin{center}
\includegraphics[width=3in,height=5in,angle=90]{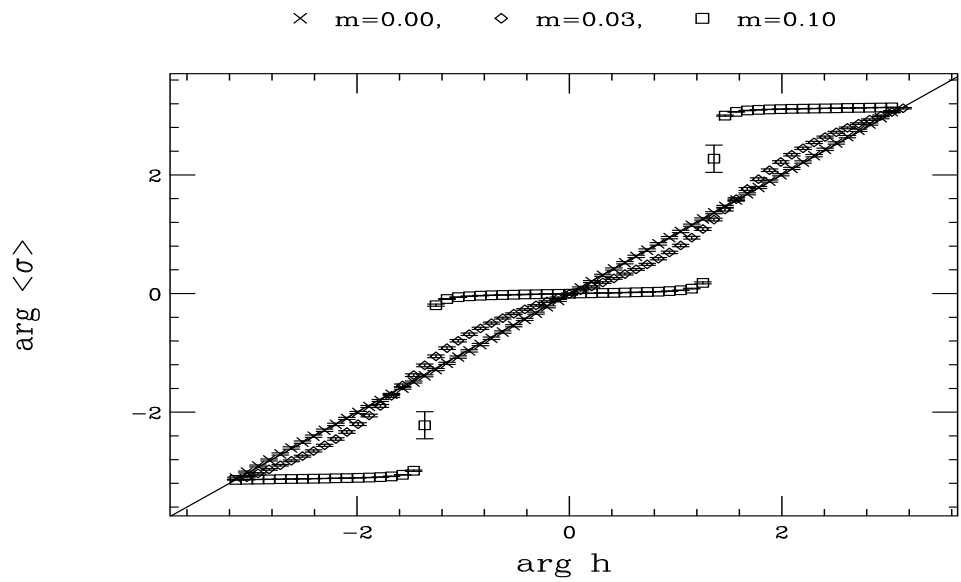}
\end{center}
\caption{A test of $U(1)_R$ symmetry,
by comparison to the prediction $\arg \vev{\s}_h = \arg h$.}
\label{argcomp}
\end{figure}

In Fig.~\ref{argcomp}
we display $\arg \vev{\s}_h$ versus $\arg h$
at $(g,|h|,N)=(0.03,0.001,16)$ for three
different mass values, $m=0,0.03,0.10$.
For $m=0$, the data passes through the (diagonal) straight
line $\arg \vev{\s} = \arg h$, showing that $U(1)_R$ is a
symmetry of the effective potential.  For
$m=0.03$, the data deviates slightly from
the straight line, indicating that
the symmetry is only slightly violated,
breaking to $Z_2(R)$.
Finally, at $m=0.10$, the $U(1)_R$ symmetry is completely
broken.  The fact that $\arg \vev{\s}_h \approx
\pm \pi$ in this case can be understood
as follows.  For larger values of $m$ and
the very small $h$ that we choose, the potential
$V=|W'(\s)|^2$ dominates over the source
potential $\Delta V(h)$.
In that case, $\vev{\s}_h \approx \pm \vev{\s}_{0^+}
\equiv \pm v$.  For $m,g$ positive, $v>0$.
The role of $h$ then is just as a perturbation
to pick the sign of $\pm v$.
It follows that $\arg \vev{\s}_h \approx 0, \pm \pi$.

\begin{figure}
\begin{center}
\includegraphics[width=3in,height=5in,angle=90]{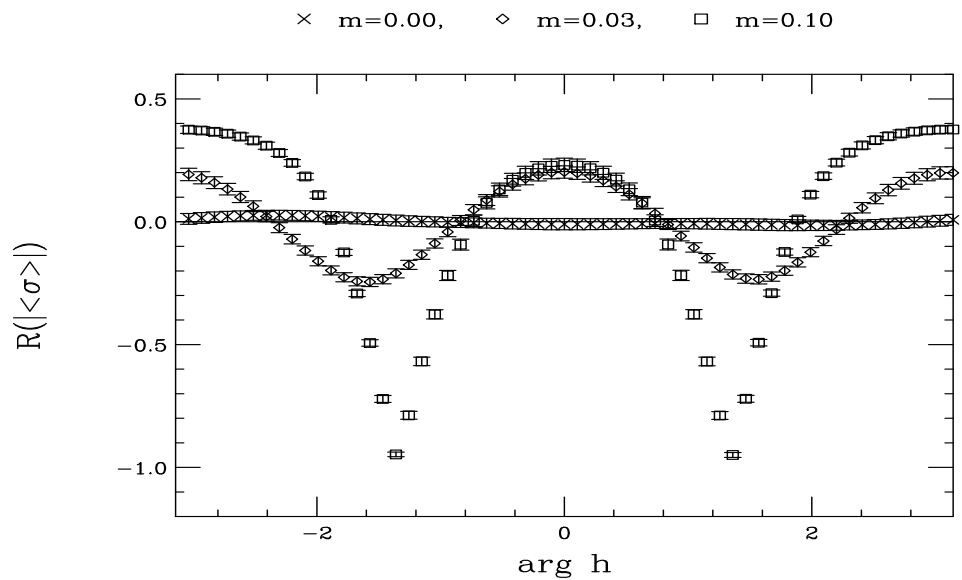}
\end{center}
\caption{
A test of $U(1)_R$ symmetry, by comparison to the 
prediction that $|\vev{\s}_h|$
should be independent of $\arg h$.  Relative deviation
from the average w.r.t.~$\arg h$ is measured by $R$.}
\label{rmagcomp}
\end{figure}

The second part of the conjecture \myref{u1con}
was studied through the quantity
\beq
R(|\vev{\s}_h|) = \frac{|\vev{\s}_h|-\overline{|\vev{\s}|}}
{\overline{|\vev{\s}|}}, \quad 
{\overline{|\vev{\s}|}} = \frac{1}{n} \sum_{j=1}^n |\vev{\s}_{h_j}|
\label{Rdf}
\eeq
where $h_j=|h| \exp(2 \pi i j/n)$ corresponds to
the values of $h$ that were used in the data set.
$R$ measures the relative shift of $|\vev{\s}_h|$ away
from the mean w.r.t.~$\arg h$.
In Fig.~\ref{rmagcomp}, one sees that $U(1)_R$ is restored
at $m=0$.

\bfe{Scaling.}
In research in progress,
we are performing another nonperturbative check
of the lattice.  As mentioned
at the outset, the continuum
theory in the critical domain is believed
to afford a Landau-Ginzburg description
of the minimal discrete series of $\Ncal=2$
superconformal field theories; the critical exponents of
relevant operators are known exactly.
The lattice theory should
reproduce these exponents.  We are
studying this through an
examination of hyperscaling (dependence
on correlation length) and finite-size 
scaling (dependence on system size).
We hope to
report the results of that study in the
near future.  Unfortunately, on the
large lattice sizes required for such an analysis,
a variety numerical obstacles have been
found to arise in our simulations.
Due to space limitations, we do not
detail them here.

\bfe{Interpretation.}
The simulation results related to R-symmetries
are quite encouraging.  The explicit breaking
due to the Wilson mass term in the superpotential is harmless
in the continuum limit; the continuum R-symmetry
is recovered without the need for counterterms.

Undoubtedly these positive results are related
to the following features: (i) the symmetry breaking is due
to irrelevant operators; (ii) 1PI diagrams
of UV degree $D \geq 0$ do not occur in 
the lattice perturbation series.
The cancellations of $D=0$ contributions of
subdiagrams in lattice perturbation theory
is intimately related to the exact lattice 
supersymmetry \cite{Giedt:2004qs}.
It would be very interesting to know whether or
not other lattice actions with an exact
supersymmetry, such as the super-Yang-Mills
examples that have been recently proposed
\cite{Cohen:2003xe}, 
have a finite lattice perturbation series,
in the sense that they have no $D \geq 0$
1PI diagrams.  However,
a careful power-counting analysis, comparable
to that done by Reisz for 4d Yang-Mills \cite{Reisz:1988kk}, has yet to
be performed.

\end{document}